# Albite feldspar dissolution kinetics as a function of the Gibbs free energy at high pCO₂

R. Hellmann
*Environmental Geochemistry Group, LGIT; Univ. Grenoble, CNRS UMR C5559, OSUG- Grenoble, France*

D. Daval
*Laboratoire de Géologie, Ecole Normale Supérieure de Paris, CNRS UMR 8538- Paris, France*

D. Tisserand & F. Renard
*LGIT; Univ. Grenoble, CNRS UMR C5559, OSUG- Grenoble, France*

ABSTRACT: We are currently measuring the dissolution kinetics of albite feldspar at 100 °C in the presence of high levels of dissolved $CO_2$ ($pCO_2$ = 9 MPa) as a function of the saturation state of the feldspar (Gibbs free energy of reaction, $\Delta G$). The experiments are conducted using a flow through reactor, thereby allowing the dissolution reactions to occur at a fixed pH and at constant, but variable saturation states. Preliminary results indicate that at far-from-equilibrium conditions, the dissolution kinetics of albite are defined by a rate plateau, with $R \approx 5.0 \times 10^{-10}$ mol m$^{-2}$ s$^{-1}$ at $-70 < \Delta G < -40$ kJ mol$^{-1}$. At $\Delta G > -40$ kJ mol$^{-1}$, the rates decrease sharply, revealing a strong inverse relation between the dissolution rate and free energy. Based on the experiments carried out to date, the dissolution rate-free energy data correspond to a highly non-linear and sigmoidal relation, in accord with recent studies.

## 1 INTRODUCTION

Increasing environmental and societal concerns about rising levels of $CO_2$ in the atmosphere, due mostly to anthropogenic sources, has led to the development of various strategies for combating the inexorable rise of $CO_2$. Among the considered strategies is carbon capture and storage (CCS), which is based on the injection of massive amounts of $CO_2$ into underground repositories, such as aquifers and sedimentary basins (see review in Schiermeier, 2006). Injection of $CO_2$ underground generally results in a substantial pH decrease and acidification of the pore waters (e.g. Kharaka et al., 2006), due to the production of carbonic acid. As a consequence, mineral-water interactions, which are very pH-sensitive, may play a substantial role in determining the mechano-chemical stability of underground CCS sites. With this in mind, scientists require accurate experimental data concerning the reactivity of minerals subject to interaction with low pH pore fluids, which are crucial for accurately predicting the suitability and performance of planned CCS sites.

Past studies have shown that silicate mineral-weathering is more effective at removing atmospheric $CO_2$ than carbonate mineral weathering (Berner et al., 1983; Berner & Lasaga, 1989). Taking the following equation to represent carbonate weathering:

$$CaCO_3 + CO_2 + H_2O \Leftrightarrow Ca^{2+} + 2HCO_3^- \quad (1)$$
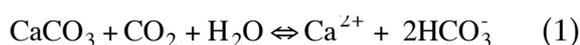

when considering the dissolution reaction (left to right), one mole of $CO_2$ is consumed for one mole of $Ca^{2+}$ and two moles of $HCO_3^-$ produced. However, the back reaction (right to left), resulting in the precipitation of calcium carbonate, releases one mole of $CO_2$, and thus there is no net consumption of $CO_2$. The weathering of Ca or Mg silicates can be represented with this equation:

$$Ca(Mg)SiO_3 + 2CO_2 + 3H_2O \Leftrightarrow Ca^{2+} + 2HCO_3^- + H_4SiO_4 \quad (2)$$
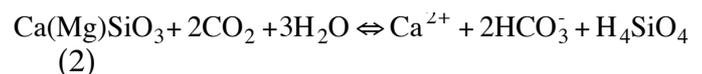

In the reaction above, two moles of $CO_2$ are consumed during dissolution, but in this case, the back reaction will result in precipitation of calcium carbonate, leading to the consumption of one mole of $CO_2$. Therefore, the coupled reactions of Ca and Mg silicate dissolution and reprecipitation of calcium or magnesium carbonate phases results in the net consumption of $CO_2$ (see e.g. Toutain et al., 2002). Based on the difference in $CO_2$ consumption between equations 1 and 2, the advantage of injecting massive amounts of $CO_2$ into CCS sites containing mixed oxide silicate minerals is apparent.

The dissolution kinetics of feldspars and other mixed oxide silicates in the presence of high $pCO_2$ aqueous fluids have been examined in a few previous experimental studies (e.g. Lagache, 1965, 1976; Brady & Carroll, 1994; Berg & Banwart, 2000; Golubev et al., 2004; Carroll and Knauss, 2005). In general, these studies showed that the effect of $CO_2$ on the dissolution kinetics is primarily due to acidifi-cation, and that the intrinsic effect of $CO_2$ is at most only of minor importance. Moreover, a number of limited studies have recently addressed the coupling of chemical reactions with the mechanical response of host rocks subject to the injection of $CO_2$, based on experimental (Le Guen et al., submitted) and theoretical approaches (Renard et al., 2005).

In this study, we are investigating the dissolution behavior of a feldspar in the presence of water containing elevated $CO_2$ levels at 100 °C and 9 MPa fluid pressure. These *P,T* conditions represent shallow $CO_2$ injection and sequestration conditions (< 1 km). Because the kinetics are measured as a function of saturation state, the results from this study are applicable to a wide range of conditions, ranging from reactions at far from equilibrium close to the injection well, to near-equilibrium conditions that are representative of far-traveled $CO_2$ plumes.

The choice of investigating a non-Ca bearing feldspar (albite) may seem counterintuitive (see Eq. 2), given that $CO_2$ sequestration most commonly involves precipitation of Ca, Mg, or Fe carbonate phases. However, lesser known carbonate phases, such as dawsonite, $NaAlCO_3(OH)_2$, may also precipitate (Knauss et al., 2005) and play an important role in $CO_2$ sequestration. In addition, Ca and Mg are often present at important concentrations in most aqueous fluids, and their reaction with $HCO_3^-$ produced by $CO_2$ dissolution in water can also lead to carbonate precipitation and $CO_2$ sequestration. A final reason for using albite feldspar in these experiments is that the dissolution kinetics of this phase are better documented than for the other feldspars, and this provides an important means for comparison of data bases.

## 2 METHODS

The experiments are conducted using albite feldspar ($NaAlSi_3O_8$) with a granulometry of 0.5-0.8 mm. A precision high pressure liquid chromatography pump injects pure water into a static, 1 L autoclave containing two horizontally-stratified phases: supercritical $CO_2$ (on top) and water (bottom) at ambient *T* and *P* = 9.0 MPa. Dissolution of the $CO_2$ into the water serves to acidify it to an estimated pH of 3.0 based on calculations using the EQ3NR code (Wolery, 1992). Using estimates of the fugacity coefficient from empirical correlation diagrams (Fig. 9, Spycher et al., 2003), the $fCO_2$ was estimated to equal 7.2 MPa. The continuous pumping of fresh water into the static autoclave forces the $CO_2$-rich water to flow into a 300 mL continuously stirred reactor that contains the albite grains. The fluid pressure in the stirred reactor also equals 9.0 MPa, which is set by a back pressure regulator. Fluid samples are continuously collected and the concentrations of Si, Al, and Na measured by ICP using matrix-matched standards. Silicon is also measured by silicomolybdate colorimetry.

The rates of dissolution are based on the steady-state effluent concentrations of Si, Al, and Na that are normalized with respect to albite stoichiometry. To date, only a single, continuous experiment has been run, but at variable saturation states. The experiment started at far-from-equilibrium conditions with a flow rate of 0.1 mL min$^{-1}$ (reactor residence time of ~2.1 d). For each given flow rate, steady-state dissolution conditions are ensured by allowing the reactor concentrations to attain constant values over periods of several weeks, and as conditions closer to equilibrium are imposed by decreasing the flow rate, the periods necessary for attainment of steady state are correspondingly longer, on the order of 1-2 months. At present the dissolution experiment is being run with a flow rate of 0.004 mL min$^{-1}$ (residence time ~52 d). The last experimental data set, closest to equilibrium, will be collected at 0.001 mL min$^{-1}$. The $\Delta G$ value for each flow rate is calculated using the steady-state concentrations of Si, Al and Na and the EQ3NR code (for details on calculations, see Hellmann & Tisserand, 2006). The theoretically (EQ3NR) determined pH of pure water enriched in $CO_2$ (log fugacity$_{CO2}$ = 1.86) at 100 °C inside the reaction vessel is 3.25, the ionic strength = 0.000578.

## 3 RESULTS

At conditions furthest from equilibrium, the dissolution rates are approximately constant and independent of the free energy, $\Delta G$, and therefore, independent of the aqueous concentrations of Si, Al and Na. The constant rates can be characterized in terms of a dissolution rate plateau, with a value of $R \approx 5.0 \times 10^{-10}$ mol m$^{-2}$ s$^{-1}$ over a saturation range of $-70 < \Delta G < -40$ kJ mol$^{-1}$. This rate is in good agreement with far-from-equilibrium rates that can be extrapolated from the albite kinetic measurements by Hellmann (1994). In the present study, the individual rates based on Si, Al, and Na are roughly equal, indicating that dissolution is stoichiometric at far-from-equilibrium conditions. At $\Delta G > -40$ kJ mol$^{-1}$, the measured rates decrease sharply with increasing saturation, indicating an inverse relation between the kinetics of dissolution and the free energy. At present, the highest $\Delta G$ attained equals $-34$ kJ mol$^{-1}$.

Over the range $\Delta G > -40$ kJ mol$^{-1}$, the dissolution behavior becomes increasingly non-stoichiometric, with $R_{Si} \approx R_{Na} \gg R_{Al}$. Even though the reacted albite has not yet been examined, the non-stoichiometry of the rates most probably results from the precipitation of secondary oxy-hydroxide Al phase(s). This possibility is supported by theoretical predictions of supersaturation of several Al-phases based on EQ3NR calculations. In addition, there is evidence at

$\Delta G$ = -34 kJ mol$^{-1}$ that $R_{Si} > R_{Na}$. It will be interesting to determine whether this small difference in rates is amplified at conditions closer to equilibrium.

Even though the experimental results are still incomplete with respect to conditions closer to equilibrium, it is possible to directly compare the present results with those from a similar study using albite feldspar (Hellmann & Tisserand, 2006). In the latter study, the experimental $R$-$\Delta G$ relation was determined to be sigmoidal and highly non-linear, based on dissolution at 150 °C and basic pH (9.2). Therein the $R$-$\Delta G$ relation was decomposed into 3 distinct regions: a rate plateau at far from equilibrium, a 'transition equilibrium' region characterized by a sharp decrease in dissolution rates with increasing $\Delta G$, and a near-equilibrium region where the rates decrease towards equilibrium, but with a much weaker dependence on $\Delta G$. When the rates from the present study are normalized with respect to the plateau rate of 5.0 x 10$^{-10}$ mol m$^{-2}$ s$^{-1}$, it is possible to directly compare these data with those from Hellmann & Tisserand (2006). As evidenced in Fig. 1, the agreement between the two data trends is relatively good with respect to the first two regions of the sigmoidal $R$-$\Delta G$ relation. Since our present study is still in progress, further data will allow us to test whether this agreement holds for saturations states closer to equilibrium.

One important aspect that we have not yet been able to address is how the rates are affected by the temporal evolution of the total surface area. This parameter depends on both the sample mass and the specific surface area, and both of these change with time. It is possible that some of the discordance between the present data and those reported in Hellmann & Tisserand (2006) can be decreased by adjustment of the total surface area, and thereby changing the rates.

As mentioned in the Methods section, the calculated in situ pH = 3.25 in the absence of albite. The in situ pH rose to 3.28 during dissolution at the lowest degree of saturation ($\Delta G \approx$ -70 kJ mol$^{-1}$), and equaled 3.38 at $\Delta G$ = -34 kJ mol$^{-1}$. This follows as a consequence of our not choosing to use a pH buffer in order to avoid any potential rate-influencing effects. The pH increase, which up to this point is not important, may become more important as conditions closer to equilibrium are approached.

Even though our $R$-$\Delta G$ data set is still incomplete, the present study suggests that the $R$-$\Delta G$ relation is sigmoidal and highly non-linear, which is in agreement with other studies on feldspars (Hellmann & Tisserand, 2006; Beig & Lüttge, 2006; Burch et al., 1993) and a few other silicate minerals (e.g. Nagy & Lasaga, 1992). Thus, one possibility for representing an overall rate law that integrates the effect of $\Delta G$ is via the following expression, originally based on BCF theory and developed by Burch et al. (1993):

$$R = k_1 [1 - \exp(-ng^{m1})] + k_2 [1 - \exp(-g)]^{m2} \quad (3)$$

In the equation above $k_1$ and $k_2$ are rate constants (units mol m$^{-2}$ s$^{-1}$) determined by regression, $g \equiv |\Delta G_r|/RT$ is a dimensionless number (R is the gas constant, $T$ is in K), and $n$, $m_1$, and $m_2$ are adjustable fitted parameters. The above expression is actually based on two dissolution mechanisms operating simultaneously, and it is a critical $\Delta G$ value that delimits the free energy range in which each mechanism dominates (see details in Burch et al., 1993; see also discussion and further references in Hellmann & Tisserand, 2006).

Of equal importance is the fact that the $R$-$\Delta G$ data in the present study are not at all in agreement with the 'TST' rate expression shown in Fig. 1. The TST rate expression is based on the following function for the free energy:

$$f(\Delta G_r) = \left[1 - \exp\left(n \frac{\Delta G_r}{RT}\right)\right] \quad (4)$$

As shown in Fig. 1, the TST rate equation considerably over-estimates the dissolution rate when $\Delta G$ > -40 kJ mol$^{-1}$. Moreover, the use of TST rate equations in geochemical codes is wide spread, and its use for modeling dissolution associated with $CO_2$ injection into underground aquifers would lead to significant errors in the prediction of dissolution rates of feldspar, as well as the rate of $CO_2$ consumption. For this reason, one of the most important goals of the present study is to provide the geochemical community with an overall rate law for feldspars that couples kinetics with saturation state. This will allow for more accurate modeling of feldspar-water interactions associated with the underground migration of a $CO_2$ plume, starting from the injection well and extending to far-field environments where chemical conditions are close to equilibrium.

4 CONCLUSIONS

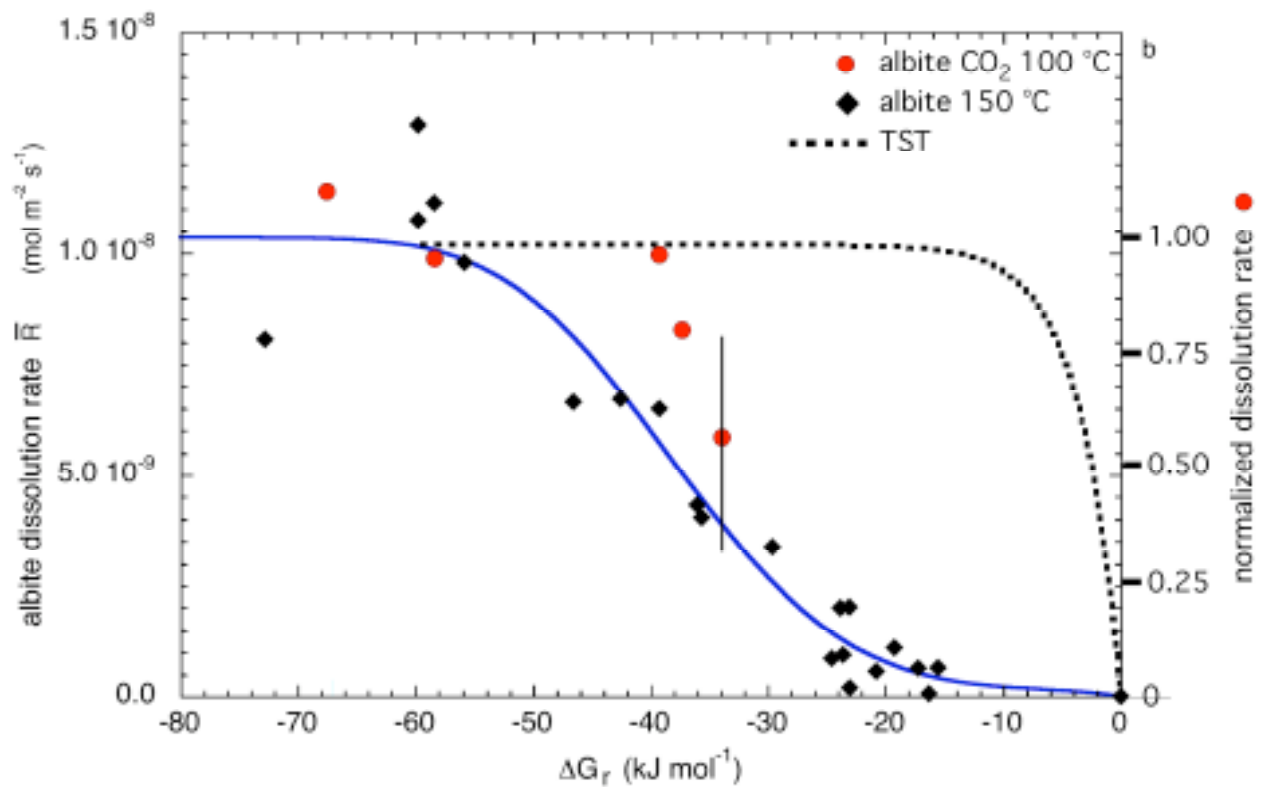

Figure 1. Dissolution rates ($R$) as a function of free energy, $\Delta G$. Filled circles represent data from present study (right $y$ axis), filled diamonds and curve (left $y$ axis) are from a study of albite dissolution at 150 °C by Hellmann & Tisserand (2006). The $R$-$\Delta G$ trends from both data sets can be qualitatively compared by equating the normalized plateau rates in each study (i.e. $R_{plateau}$ = 1). The Hellmann & Tisserand (2006) rate plateau equals $1.0 \times 10^{-8}$ (which can be equated to $R = 1$ when normalized); the rates in the present study have been normalized by dividing each rate by $5.0 \times 10^{-10}$, where $R = 5.0 \times 10^{-10}$ represents the plateau dissolution rate.

ACKNOWLEDGEMENTS

This research has been financed by Gaz de France contract # 722478/00 (program coordinators: Christophe Rigollet, Samuel Saysset, Rémi Dreux).

REFERENCES

Berg, A. & Banwart, S.A. 2000. Carbon dioxide mediated dissolution of Ca-feldspar: implications for silicate weathering. *Chem. Geol.* 163: 25-42.

Berner, R.A. & Lasaga, A.C. 1989. Modeling the geochemical carbon cycle. *Sci. Am.* 260: 74-81.

Berner, R.A., Lasaga, A.C., & Garrels, R.M. 1983. The carbonate-silicate geochemical cycle and its effects on atmospheric carbon dioxide over the past 100 million years. *Am. J. Sci.* 284: 1183-1192.

Beig, M.S. & Lüttge, A. 2006. Albite dissolution kinetics as a function of distance from equilibrium: Implications for natural feldspar weathering. *Geochim. Cosmochim. Acta* 70: 1402-1420.

Brady, P.V. & Carroll, S.A. 1994. Direct effects of $CO_2$ and temperature on silicate weathering: Possible implications for climate control. *Geochim. Cosmochim. Acta* 58: 1853-1856.

Burch, T.E., Nagy, K.L., & Lasaga, A.C. 1993. Free energy dependence of albite dissolution kinetics at 80°C and pH 8.8. *Chem. Geol.* 105: 137-162.

Carroll, S.A. & Knauss, K.G. 2005. Dependence of labradorite dissolution kinetics on $CO_{2(aq)}$, $Al_{(aq)}$, and temperature. *Chem. Geol.* 217: 213-225.

Golubev, S.V., Pokrovsky, O.S., & Schott, J. 2004. Laboratory weathering of Ca- and Mg-bearing silicates: weak effect of $CO_2$ and organic ligands. *Geochim. Cosmochim. Acta*, A418.

Hellmann, R. 1994. The albite-water system: Part I. The kinetics of dissolution as a function of pH at 100, 200, and 300°C. *Geochim. Cosmochim. Acta* 58: 595-611.

Hellmann, R. & Tisserand, D. 2006. Dissolution kinetics as a function of the Gibbs free energy of reaction: An experimental study based on albite feldspar. *Geochim. Cosmochim. Acta* 70: 364-383.

Kharaka, Y.K., Cole D.R., Hovorka, S.D., Gunter, W.D., Knauss, K.G. & Freifeld, B.M. 2006. Gas-water interactions in Frio Formation following $CO_2$ injection: Implications for the storage of greenhouse gases in sedimentary basins. *Geology* 34: 577-580.

Knauss, K.G., Johnson, J.W., & Steefel, C.I. 2005. Evaluation of the impact of $CO_2$, co-cotaminant gas, aqueous fluid and resevoir rock interactions on the geologic sequestration of $CO_2$. *Chem. Geol.* 217: 339-350.

Lagache, M. 1965. Contribution à l'étude de l'altération des feldspaths, dans l'eau, entre 100 et 200°C, sous diverses pressions de $CO_2$, et application à la synthèse des minéraux argileux. *Bull. Soc. franç. Minér. Crist.* 88: 223-253.

Lagache, M. 1976. New data on the kinetics of the dissolution of alkali feldpars at 200°C in $CO_2$ charged water. *Geochim. Cosmochim. Acta* 40: 157-161.

Le Guen, Y., Renard, F., Hellmann, R., Brosse, E., Collombet, M., Tisserand, D., & Gratier, J.P. (in revision) Enhanced deformation of limestone and sandstone in the presence of high $pCO2$ fluids. *J. Geophys. Res*.

Nagy, K.L. & Lasaga, A. C. 1992. Dissolution and precipitation kinetics of gibbsite at 80°C and pH 3: the dependence on solution saturation state. *Geochim. Cosmochim. Acta* 56: 3093-3111.

Renard, F., Gundersen, E., Hellmann, R. Collombet, M., & Le Guen, Y. 2005. Numerical modeling of the effect of carbon dioxide sequestration on the rate of pressure solution creep in limestone: preliminary results. *Oil & Gas Science and Technology - Rev. IFP* 60: 381-399.

Schiermeier, Q. 2006. Putting the carbon back. *Nature* 442: 620-623.

Spycher, N., Pruess, K., & Ennis-King, J. 2003. $CO_2$-$H_2O$ mixtures in the geological sequestration of $CO_2$. I. Assessment and calculation of mutual solubilities from 12 to 100 °C and up to 600 bar. *Geochim. Cosmochim. Acta* 67: 3015-3031.

Toutain, J-P., Baubron, J.-C. & François L. 2002. Runoff control of soil degassing at an active volcano. The case of Piton de la Fournaise, Réunion Island. *Earth Planet. Sci. Lett*. 197: 83-94.

Wolery, T.J. 1992. *EQ3NR, A Computer Program for Geochemical Aqueous Speciation-Solubility Calculations: Theoretical Manual, User's Guide, and Related Documentation (Version 7.0)*. Lawrence Livermore Natl. Lab. UCRL-MA-110662 PT III.